# Observation of topological valley transport of sound in sonic crystals

Jiuyang Lu[1,4], Chunyin Qiu[1*], Liping Ye[1], Xiying Fan[1], Manzhu Ke[1],
Fan Zhang[2], and Zhengyou Liu[1,3*]

[1]Key Laboratory of Artificial Micro- and Nano-structures of Ministry of Education and School of Physics and Technology, Wuhan University, Wuhan 430072, China
[2]Department of Physics, University of Texas at Dallas, Richardson, Texas 75080, USA
[3]Institute for Advanced Studies, Wuhan University, Wuhan 430072, China
[4]Department of Physics, South China University of Technology, Guangzhou 510641, China

**Valley pseudospin, labeling quantum states of energy extrema in momentum space, is attracting tremendous attention[1-13] because of its potential in constructing new carrier of information. Compared with the non-topological bulk valley transport realized soon after predictions[1-5], the topological valley transport in domain walls[6-13] is extremely challenging owing to the inter-valley scattering inevitably induced by atomic scale imperfectness, until the recent electronic signature observed in bilayer graphene[12,13]. Here we report the first experimental observation of topological valley transport of sound in sonic crystals. The macroscopic nature of sonic crystals permits the flexible and accurate design of domain walls. In addition to a direct visualization of the valley-selective edge modes through spatial scanning of sound field, reflection immunity is observed in sharply curved interfaces. The topologically protected interface transport of sound, strikingly different from that in traditional sound waveguides[14,15], may serve as the basis of designing devices with unconventional functions.**

The macroscopic controllability enables the sonic crystals (SCs) to be tractable classical counterparts for exploring the complex quantum physics requiring atomic scale manipulations. Two-dimensional (2D) hexagonal crystals, such as graphene, bilayer graphene, and transition-metal dichalcogenides, exhibit a pair of degenerate states at the inequivalent $K$ and $K'$ valleys in momentum space. Inversion symmetry breaking generically gaps the degeneracy and gives rise to a tantalizing valley Hall insulator[1-13], in which valley-chiral electrons counterpropagate at the boundary in the absence of inter-valley scattering. To observe the topological valley transport in the acoustic system, an experimental knob is required to trigger the valley Hall phase transition. Instead of breaking the inversion symmetry like in graphene systems[7-13], here we introduce a mirror symmetry breaking mechanism to induce the acoustic analog of the topological semimetal-insulator transition. To do this, anisotropic scatterers, which have been employed to optimize the performance of the SCs[16,17], are used to close and reopen the band gaps[18,19]. Interestingly, by simply rotating the anisotropic scatterers, we can control the frequency gap of the acoustic insulators and the shape of the phase domain wall flexibly.

As depicted in Fig. 1a, the SC consists of a triangular-lattice array (with lattice constant 4.3 cm) of rod-like scatterers placed in a 2D air waveguide formed by two parallel plastic plates (Methods). Here the top plate is removed for visualization. The rod has a shape of regular triangle (with side length 3.0 cm), and its orientation degree of freedom, characterized by the rotation angle $\alpha$, enables the system to exhibit different symmetries. For $\alpha = m\pi/3$, with $m$ being an integer, the point group at $K$ (and $K'$) is featured by $C_{3v}$ symmetry, due to the perfect match of the mirrors of individual scatterers to those of the triangular-lattice. Apart from the specific angles, the mirrors are mismatched and the symmetry reduces to $C_3$. Therefore, a twofold Dirac degeneracy is protected at $K$ (and $K'$) for any SC with $\alpha = m\pi/3$, whereas the degeneracy would be lifted for any other rod orientation because of the broken mirror symmetry. This is exemplified by the dispersions for the SCs with $\alpha = 0^o$ and $-10^\circ$ (Fig. 1b). To reveal the characteristics of the Dirac degeneracy, Fig. 1c presents the field patterns for the two degenerate states at $K$. Both states, labeled by $\psi_{p^-}^0$ and $\psi_{q^+}^0$, display typical vortex profiles centered at the two inequivalent triangular-lattice centers $p$ and $q$, where the sign $\pm$ indicates an anticlockwise or clockwise energy flow. The vortex chirality can be viewed as a pseudospin analogous to the A-B sublattice or the top-bottom layer pseudospin in graphene systems[6-8]. The counterparts at valley $K'$, denoted by $\psi_{p^+}^0$ and $\psi_{q^-}^0$, possess invariant vortex cores but opposite chirality, as required by time-reversal symmetry. Below we focus on the physics at valley $K$, and that at $K'$ can be derived directly from time-reversal symmetry.

The acoustic valley Hall (AVH) phase transition has been convincingly observed in our simulations and experiments. Figure 1d provides a continuous evolution of the band edge frequencies $\omega_{p^-}$ and $\omega_{q^+}$ versus the rotation angle $\alpha$. Evidently, the frequency order of the first two bands, locked to the opposite vortex pseudospins, is inverted at $\alpha = 0^\circ$, a signal of the AVH phase transition. This is analogous to the semimetal-insulator transition induced by an electric field in bilayer graphene[7,8]. We have measured the band-edge frequencies for the SCs with different $\alpha$. The experimental data in Fig. 1d (circles) confirm the gap closure and reopening as the scatterer is rotated through $\alpha = 0^\circ$. To further identify different AVH insulators, the phase profiles of the $K$ valley states are scanned around the phase singular point $p$ or $q$. As exemplified by the SCs with $\alpha = \pm 10^\circ$ (Fig. 1e), the overall slopes of the experimentally measured phase distributions (circles) agree well with those extracted from the eigenfields (lines). Intriguingly, the vortex states labeled by $p^-$ and $q^+$ carry quantized angular momenta $-1$ and $+1$, respectively, owing to the threefold rotation symmetry of the SC.

The AVH phase transition can be captured by an $\alpha$-dependent continuum

Hamiltonian, $\delta H = v_D \delta k_x \sigma_x + v_D \delta k_y \sigma_y + m v_D^2 \sigma_z$, derived from the $\mathbf{k} \cdot \mathbf{p}$ perturbation method and spanned by the degenerate vortex pseudospins $\psi_{p^-}^0$ and $\psi_{q^+}^0$ (Methods). Here $v_D$ is the Dirac velocity of the conic dispersion at $\alpha = 0°$, $\delta \mathbf{k}$ is the momentum deviation from the $K$ point, and $\sigma_i$ are Pauli matrices of the vortex pseudospins. Consistent with the above band inversion picture (Fig. 1d), the sign of the effective mass $m = \left(\omega_{q^+} - \omega_{p^-}\right)/2v_D^2$ characterizes two different AVH insulators separated by the Dirac semimetal phase with $m = 0$ in the phase diagram. Similarly to the case of graphene, the massive Dirac Hamiltonian $\delta H$ produces a nontrivial Berry curvature $\Omega(\delta \mathbf{k}) = \frac{1}{2} m v_D \left(\delta k^2 + m^2 v_D^2\right)^{-3/2}$ [2,8] in the first band, which can be integrated into a topological charge $C_K = \mathrm{sgn}(m)/2$ [6-11]. Therefore, for a SC interface separating distinct AVH insulators, the difference in the topological charge across the interface is quantized, i.e., $|\Delta C_K| = 1$, which predicts a chiral edge mode propagating along the interface[6-11]. This bulk-boundary correspondence has been confirmed by analytically solving the boundary problem (Methods). For an interface orientated along an arbitrary $x'$ direction and located at $y' = 0$, the edge state has a general form $\phi = \left(c_1 \psi_{p^-}^0 + c_2 \psi_{q^+}^0\right) e^{i \delta k_{x'} x' - |m v_D y'|}$, with a linear dispersion $\delta \omega = v_D \delta k_{x'}$ centered at the projected $K$ valley onto the $x'$ direction. Owing to time-reversal symmetry, the $K'$ valley also hosts one chiral AVH edge mode, but traveling along the opposite direction.

To confirm the above picture, we first simulate the dispersion for the SC interface selected, for example, along the $x$ direction. Two different systems are studied comparatively: one is constructed by the SCs with $\alpha = 10°$ and $50°$, and the other is constructed by the SCs with $\alpha = -10°$ and $10°$ (referred to as domains I and II below). These SCs are considered intentionally since they share the same band gap between 3.82 and 4.34 kHz (Fig. 1d). As shown clearly in Fig. 2a, for the former system the edge spectrum is completely gapped since the two SCs belong to the same AVH phase; for the latter case, however, a pair of valley-chiral edge states counterpropagates at each interface (green lines) because of the mass inversion. The edge states $\phi_{I,II}^{\pm}$ and $\phi_{II,I}^{\pm}$ respectively label those gapless modes hosted by the interfaces I-II and II-I displayed in Fig. 2b, traveling along the $\pm x$ directions. The dispersions are linear near the crossing points as those predicted by the continumm model (red lines).

The presence of the topological AVH edge states has been solidly validated in experiments. As shown in Fig. 2b, a sandwich structure made of the SCs with $\alpha = -10^\circ$ (phase I) and $10^\circ$ (phase II) is used to study two different horizontal interfaces simultaneously. The sound signal is launched from a deep subwavelength-sized tube placed inside the left entrance of the SC interface, and probed by a movable microphone in the same channel. From the pressure distributions scanned separately along the interfaces I-II and II-I, we obtain the corresponding interface spectra through a Fourier transform. As shown in Fig. 2c, where the dark and bright color indicate the Fourier amplitude of low and high values, the experimental data (bright color) capture well the numerical dispersions (green lines) for the right-moving states $\phi_{I,II}^{+}$ and $\phi_{II,I}^{+}$, even in the frequency regimes beyond the bulk gap (since the bulk mode is not well-excited). The time-reversal counterparts, $\phi_{I,II}^{-}$ and $\phi_{II,I}^{-}$, have also been checked by sending sound signals from the right entrances. Note that a key feature of the topological edge state is the exponential decay of the field amplitude away from the SC interface, for which the decay length is determined only by the bulk parameter $|mv_D|$. To confirm this, we scan the pressure field along a straight line vertically traversing the interface II-I. As shown in Fig. 2d, the experimental result agrees excellently with the model prediction and full-wave simulation, which is indeed frequency insensitive (see inset).

As aforementioned, the topological AVH edge state stems inherently from the single valley physics. The meaning is twofold: each chiral edge mode (i) is projected from a specific valley and (ii) is a linear superposition of the two basis states. The inter-valley decoupling can thus lead to many fascinating transport phenomena, e.g., angularly selective excitation by external sound and negligible backscattering in sharply curved SC interfaces.

To stimulate the valley-projected interface state by a spatial Gaussian beam, an angular selection rule can be established according to the conservation of the momentum parallel to the sample boundary, $k_\parallel = k_0 \sin\gamma$. Here $k_\parallel$ is a projection of $K$ (or $K'$) on the boundary, $k_0$ is the wavevector in air space, and $\gamma$ characterizes the incident direction of the sound beam. The angular selectivity is exemplified in Fig. 3a by a vertical SC interface. As predicted, the AVH edge mode (projected from the $K$ valley) is well excited at the incident angle $\gamma = \sin^{-1}(k_\parallel / k_0) \approx -42^\circ$ (upper panel), in striking contrast to the deep suppression at $\gamma = 0^\circ$ (lower panel). To validate this phenomenon, we have experimentally measured the transmissions for a wide range of incident angles. The angular selectivity can be observed clearly in Fig. 3b (black circles), where the optimized transmission occurs in the predicted incident angle, in good agreement with the simulation (black line). Similarly, an optimal excitation emerges at $\gamma = 42^\circ$ if the positions of the two AVH phases are switched (red line and circles), and the edge mode is now contributed from the $K'$ valley. The broadening of the angular peaks stems from the presence of the exponential decay of the edge states. Remarkably, the angular selection rule is robust for the entire bulk gap, as displayed consistently by the numerical and experimental data in Fig. 3c. The horizontal SC interfaces in Fig. 2b deserve a special attention. As shown in Figs. 3d and 3e, for an incident Gaussian beam wide enough to cover both interfaces simultaneously, only the edge state in the lower channel is excited despite the fact that the upper one is also allowed by the criterion of momentum conservation. This can be understood from the parities of the edge modes (Methods).

Below we demonstrate the negligibly weak backscattering of the AVH edge mode propagating along sharply twisted SC interfaces. Such curved interfaces emerge frequently in electronic systems[12] and may cause inter-valley scattering to destroy the edge states[10]. Figure 4a shows the sound transport in a zigzag bending channel. As exemplified by the field pattern simulated at 4.06 kHz (inset), the sound travels smoothly in the curved path despite of suffering two sharp corners (bent by $120^o$). To give an exact description for the bending corners, the reflection (as revealed by the interference pattern in the inlet of the waveguide) originated from the impedance mismatch between the sample and the free space must be precluded. For this purpose, we have derived the transmission and reflection through a one-dimensional scattering matrix approach (Supplementary Information). Figure 4a shows a negligible reflection in the entire bulk gap. This behavior has not been observed previously in the sound waveguide designed by a SC[14,15]. Note that the accuracy of the scattering matrix method depends sensitively on the pressures detected (Supplementary Information). In practical experiments, we have measured the pressure in the output channel and compared it with the result for a sample containing a straight channel of the same length. As shown in Fig. 4b, the transmitted pressures for the two samples agree well in the frequency range of bulk gap, in contrast to the remarkable difference beyond the gap. This confirms the weak influence of the bending corners on the wave transport of the interface modes. More complex configurations have been further checked numerically (Supplementary Information), even for those sustaining simultaneously the symmetric and antisymmetric edge modes. Again, the unusual phenomenon is closely related to the valley-projected topological origin of the edge mode. As deduced from the theoretical model, the forward-moving modes are always projected from the same (*K*) valley, since the relative positions of the two AVH insulators keep invariant as the propagation of sound. In other words, the field profiles match well between the edge states of adjacent channels, which contribute to the preference of high transmission. (For a general defect located in the curved or straight waveguide, the inter-valley coupling could become sizable, depending on the property of the defect.) Recently, the reflection immunity has also been observed for electromagnetic waves traveling in a twisted interface separated by different photonic topological insulators[20,21]. In those systems, the gaps are produced by a magneto-electric coupling (dubbed bianisotropy) equivalent to the electronic spin-orbital interaction, which is absent in our sonic systems.

It is worth noting that, the rotating-scatterer mechanism enables easily tunable operation bandwidth and reconfigurable shape of the SC interface. These merits plus the intriguing valley transport properties, absent in the conventional SC-based waveguides[14,15], could be very useful in designing exceptional devices (e.g., for sound signal processing). Our finding may also pave the way for exploring the controllable topological phases and valley-dependent phenomena in various classical systems, which have been proved to be excellent macroscopic platforms in revealing topological properties[22-35] proposed originally in electronic systems, e.g., quantum Hall insulators[22-28], topological insulators[20,21,29-33], and topological semimetals[34,35]. Finally, the study provides a special insight on realizing topological insulators that require internal degrees of freedom, which is particularly important for the neutral scalar sound that lacks intrinsic polarization and uncouples to external fields.

**Methods**

**Modeled Hamiltonian.** It has been proved that[14], for the SC with $\alpha = 0°$ the perturbation Hamiltonian $\delta \mathbf{k} \cdot \mathbf{p}$, spanned by the degenerated states $\psi_{p^-}^0$ and $\psi_{q^+}^0$, yields conic dispersions centered at the hexagonal Brillouin zone corners due to the protection of the $C_{3v}$ symmetry, where $\mathbf{p}$ is a vectorial operator determined by the density distribution of the SC. As the scatterer is rotated, the mirror symmetry is broken, and the deterministic degeneracy is thus removed. This produces a Dirac mass term, i.e., $\delta_{ij}(\omega_i^2 - \omega_D^2)$, where $\omega_i = \omega_{p^-}$ or $\omega_{q^+}$ is the $\alpha$-dependent band-edge frequency, and $\omega_D$ is the Dirac frequency of the SC with $\alpha = 0°$. Using the detailed form of the $\mathbf{p}$ matrix, i.e., $\mathbf{p}_{11} = \mathbf{p}_{22} = 0$ and $\mathbf{p}_{21} = \mathbf{p}_{12}^* = 2\omega_D v_D (\hat{\mathbf{x}} + i\hat{\mathbf{y}})$, we obtain a compact form of the perturbation Hamiltonian $\delta H$ (see text) that satisfies the eigen-problem $\delta H \psi = \delta \omega \psi$, where $\delta \omega$ is the frequency deviation from $\omega_D$. The preciseness of the modeled Hamiltonian has been confirmed by duplicating the numerical dispersion near the Brillouin zone corner.

**Derivation of the AVH edge states.** For simplicity, we consider a SC interface orientated along the $x$ direction, formed by two SCs with $m < 0$ for $y > 0$ and $m > 0$ for $y < 0$. Substituting a trial decay solution into the eigen-problem, an edge state traveling along the $+x$ direction can be derived $\phi^+ = \left( \psi_{p^-}^0 + \psi_{q^+}^0 \right) e^{i\delta k_x x - |mv_D y|}$, together with a gapless dispersion $\delta \omega = v_D \delta k_x$ centered at the projection of $K$ onto the interface. Interestingly, because of the mirror symmetry between the two basis states (Fig. 1c), the edge state is locally symmetric with respect to a specific horizontal axis, apart from the exponential decay factor $e^{mv_D y} = e^{-|mv_D y|}$. Note that the property of the edge state is much different if the positions of the two AVH insulators are switched, since the detailed geometrical structure of the SC interface is inherently different from the original one. Now the edge state $\phi^+ = \left( \psi_{p^+}^0 - \psi_{q^-}^0 \right) e^{-i\delta k_x x - |mv_D y|}$, projected from the $K'$ valley, is locally anti-symmetric in each unit cell. This leads to the deep

suppression of the edge mode in the upper channel in Figs. 3d and 3e. Similarly, the valley-projected edge state can be derived for a SC interface with any arbitrary orientation.

**Simulations.** All full-wave simulations are accurately carried out by a commercial finite-element solver (COMSOL Multiphysics), where the triangular polymethyl methacrylate rods used in real experiments are modeled as acoustically rigid, considering the great impedance mismatch with respect to air. Note that in practical experiments, the rods (of height 1.2 cm) are closely sandwiched between two acoustically rigid parallel plates. The whole structure can be safely modeled as a 2D system, since the planar waveguide supports only the propagating mode uniform in the $z$-direction for the wavelength under consideration. To capture the weak reflection exactly from the two bent corners (Fig. 4), we must remove the influence of the multi-reflections induced by the impedance mismatch between the sample and free space. Thus we collect the forward-moving and backward-moving wave information in the input and output channels, according to the Bloch wave connection between two spatially equivalent locations in each channel. From the incoming and outgoing wave information for the corners, a simple one-dimensional scattering model (Supplementary Information) can be established as long as each phase domain is thick enough to prevent energy leakage.

**Experimental measurements.** In experiments, several hundreds of triangular polymethyl methacrylate rods fabricated by laser cutting are arranged into desired sample configurations. The sound signal is launched from a narrow tube (of diameter ~0.8 cm) and scanned by a movable microphone (of diameter ~0.7 cm, B&K Type 4187), together with an identical microphone fixed for phase reference. The sound signal is analyzed by a multi-analyzer system (B&K Type 3560B), from which both the phase and amplitude of the pressure field can be obtained. To prepare a Gaussian beam, the sound emitted from the tube is reflected by a carefully designed parabolic concave mirror, where the beam width is controlled by the mirror's size. In all measurements, absorbers are placed at the ends of the sample to reduce the unwanted reflection generated by the impedance mismatch between the sample and free space.

Geometric parameters of the experimental samples are listed below. The sample (see Fig. 2b) used to measure the dispersion is formed by three parts with identical areas, where the central one (domain II) corresponds to the SC with the rotation angle $\alpha = 10^\circ$, and the lateral two (domain I) correspond to the SC with $\alpha = -10^\circ$. Each part consists of 16x16 rods, i.e., 16 layers along the *y* direction and 16 rods for each layer along the *x* direction. The sample has a total size ~ 0.7x1.8 m$^2$. The sample used to confirm the angularly sensitive excitation of the edge modes (Figs. 3b and 3c) is made of 16 layers and 22 rods for each, and the sample involved in Fig. 3e is completely identical to that mentioned in Fig. 2b. In Fig. 4b, the sample with a zigzag (or straight) path consists of 22x40 (or 22x46) rods.

The data that support the plots within this paper and other findings of this study are available from the corresponding author upon request.

**Acknowledgements** We thank Chuanwei Zhang and Meng Xiao for fruitful discussions. This work is supported by the National Basic Research Program of China (Grant No. 2015CB755500); National Natural Science Foundation of China (Grant Nos. 11674250, 11374233, 11534013, 11574233, 11547310); Postdoctoral innovation talent support program (BX201600054). FZ was supported by the UT-Dallas research enhancement funds.

**Author contributions** C.Q. and J.L. conceived the original idea. J.L. performed the simulations. J.L., L.Y., X.F. and M.K. carried out the experiments. C.Q. and Z.L. supervised the project. C.Q., J.L., F.Z. and Z.L. analyzed the data and wrote the manuscript. All authors contributed to scientific discussions of the manuscript.

**Additional information** Supplementary information is available in the online version of the paper. Reprints and permission information are available online at www.nature.com/reprints. Correspondence and requests for materials should be addressed to C.Q. and Z. L.

**Competing Interests** The authors declare that they have no competing financial interests.

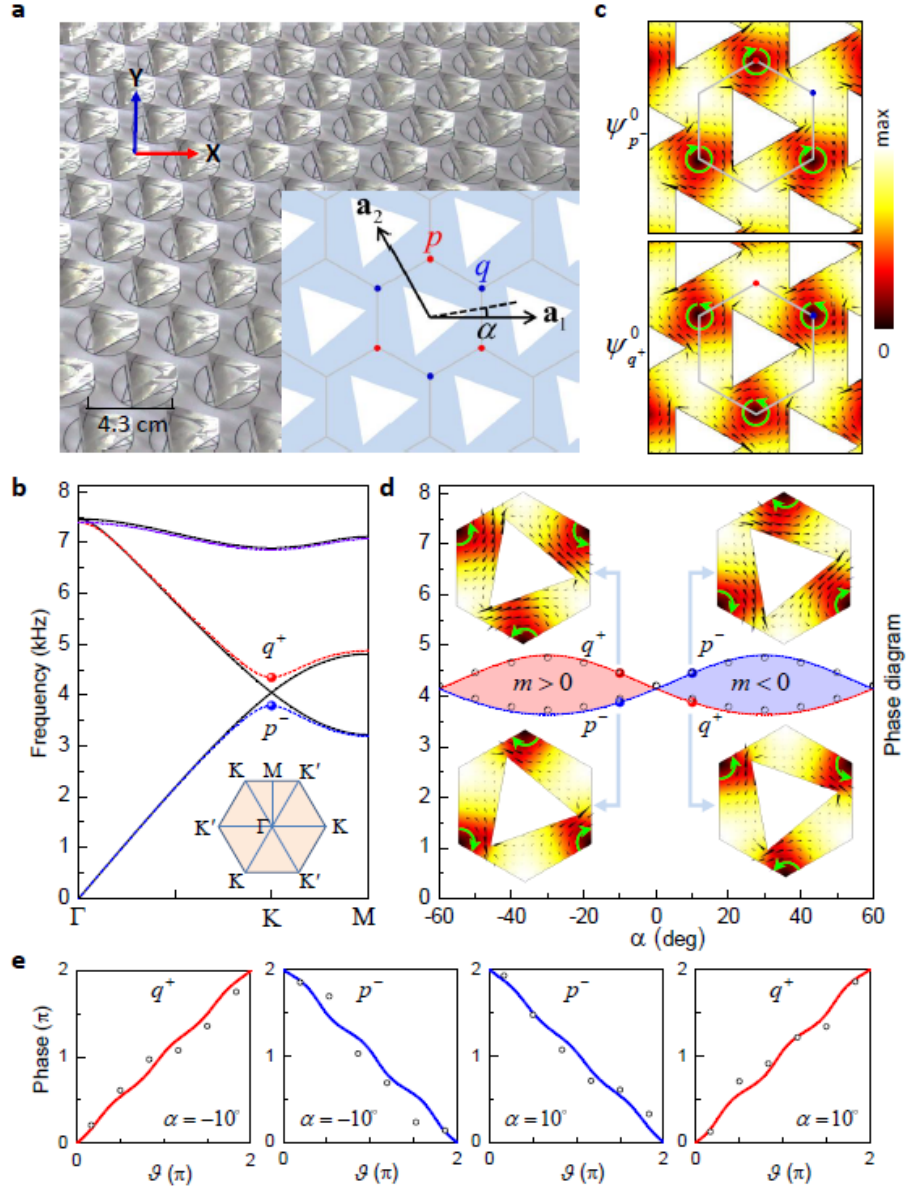

**Figure 1 Mirror symmetry breaking and topological phase transition. a**, The SC consisting of triangular polymethyl methacrylate rods positioned in a triangular-lattice with lattice vectors $\mathbf{a}_1$ and $\mathbf{a}_2$. The rotation angle $\alpha$ indicates the orientation of the scatterer with respect to $\mathbf{a}_1$. $p$ and $q$ label two inequivalent triangular-lattice centers. **b**, Gapless and gapped bulk dispersions for the cases $\alpha = 0°$ (black lines) and $-10°$ (color lines). **c**, Simulated eigenfield profiles for the degenerated $K$ states $\psi_{p^-}^0$ and $\psi_{q^+}^0$. The energy flows (arrows) manifest distinct vorticities surrounding the points $p$ and $q$ with zero pressure amplitudes (color). **d**, Phase diagram revealed by the order of band-edge frequencies (lines for simulations and circles for experiments) locked with specific vortex features (insets). Different acoustic insulating phases are characterized by their signs of the effective mass $\mathrm{sgn}(m) = \mathrm{sgn}(\omega_{q^+} - \omega_{p^-})$. **e**, Phase distributions simulated (lines) and measured (circles) anticlockwise along a circular trajectory (labeled by the azimuthal angle $\vartheta$) around the vortex cores of the $K$ valley states for the SC with $\alpha = \pm 10°$.

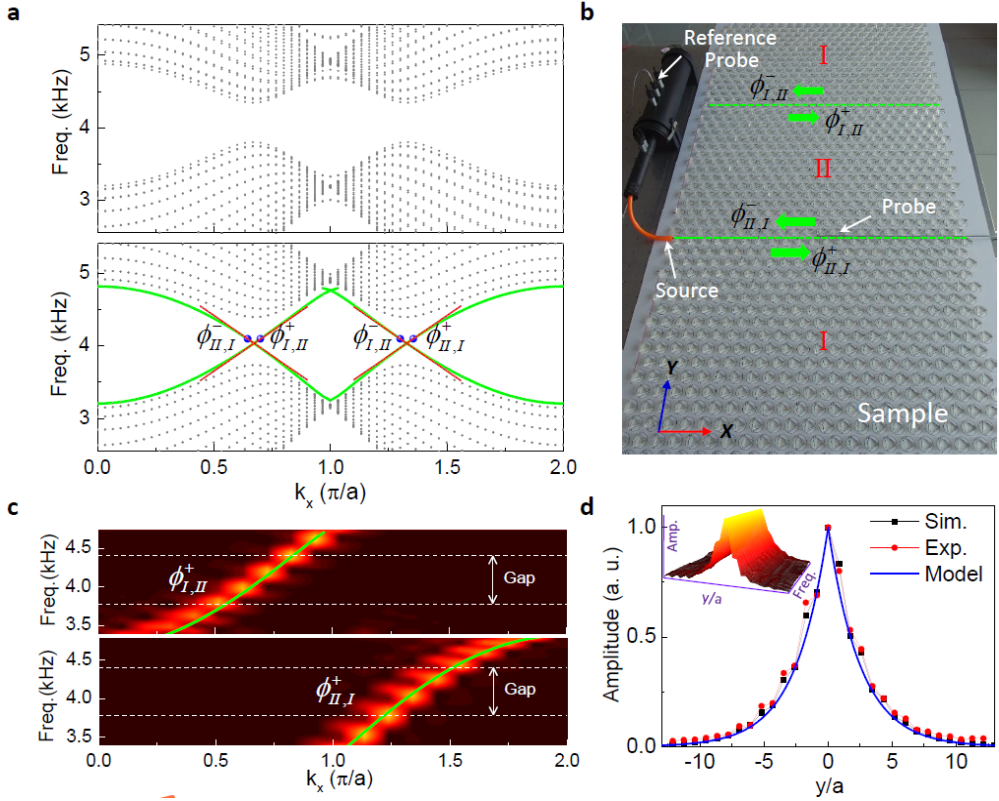

**Figure 2 Numerical and experimental validations of the topological AVH edge modes. a**, Comparative dispersions for the SC interfaces separating two topologically identical ($\alpha = 10°$ and $50°$, upper panel) and separating two topologically distinct ($\alpha = -10°$ and $10°$, lower panel) AVH insulating phases, both simulated by a superlattice structure containing two different horizontal interfaces. For the mass-inverted system, each wall supports one time-reversal pair of gapless valley-chiral edge modes (green lines), as predicted from the theoretical model (red lines). **b**, Experimental setup. The domains I and II represent the SCs with $\alpha = -10°$ and $10°$, respectively. **c**, Experimentally measured dispersions (bright color) for the edge states $\phi^+_{I,II}$ and $\phi^+_{II,I}$, compared with the numerical data (green lines). **d**,

Numerical and experimental decaying profiles for the edge state $\phi_{II,I}^{+}$ (4.06 kHz) away from the SC interface, together with the model prediction. All data are normalized by the corresponding maximum values. The inset confirms experimentally the weak frequency dependence of the exponential decay. Similar results are also observed for the edge state $\phi_{I,II}^{+}$.

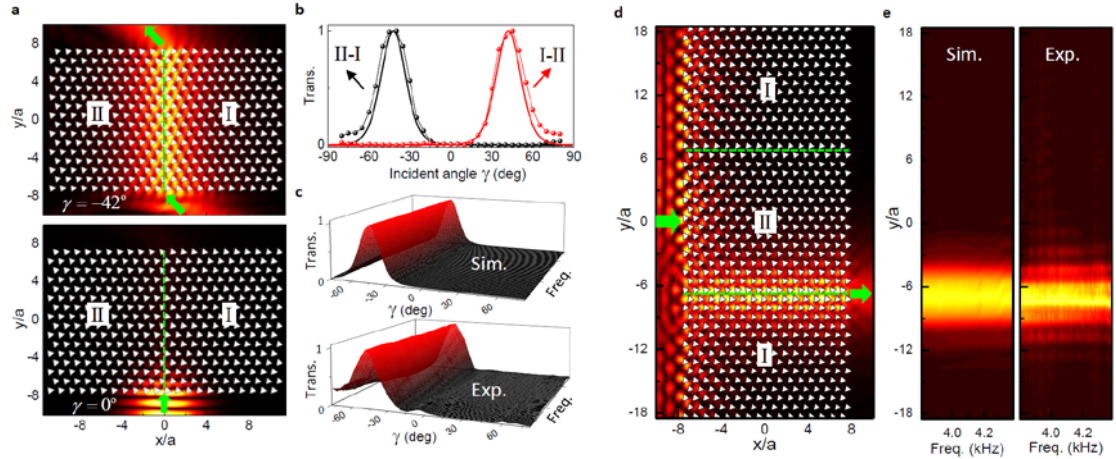

**Figure 3 Valley-selective excitation of the AVH edge mode. a**, Field distributions (4.06 kHz) simulated for the sample with a vertical SC interface II-I, excited by the Gaussian beams at the incident angles $\gamma = -42^{\circ}$ and $\gamma = 0^{\circ}$. **b**, Numerical (black line) and experimental (black circles) transmissions plotted as a function of $\gamma$, together with the data for the interface I-II (red line and circles). The transmission is normalized by the value of the angular peak. **c**, Frequency dependent angular spectra show the angular selectivity within the whole bulk gap. **d**, Amplitude distribution simulated by a wide Gaussian beam (4.06 kHz) incident normally onto the sample with two distinct horizontal interfaces. **e**, Frequency dependent pressure amplitudes scanned in the output facet of the sample, demonstrating numerically and experimentally the broadband channel selectivity according to the parity of the edge states.

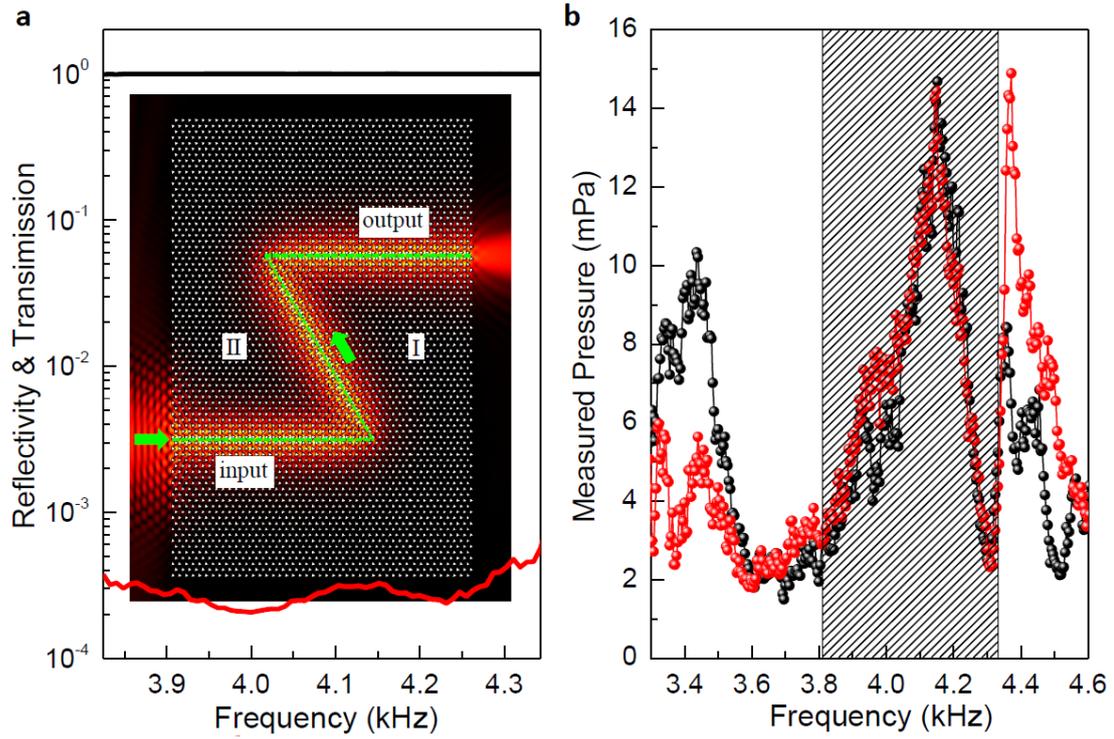

**Figure 4 Reflection immunity of the AVH edge modes from sharp corners. a**, Power transmission (black line) and reflection (red line) spectra calculated for the two sharp turns in a zigzag path. Inset: Field pattern simulated at 4.06 kHz. **b**, Transmitted pressure measured for the zigzag path, compared with that for a straight channel sample. The good agreement in the bulk gap (shadow region) shows a small influence of the bending corners to the interface transport.